\documentstyle[12pt]{article}

\def \Ref#1{{$^{#1}$}}  % Define reference style, used both in main body and
\def \Cl37{$^{37}$Cl}
\def \Ga71{$^{71}$Ga}
\def \B8{$^8$B}
\def \Be7{$^7$Be}

\begin{document}

\rightline{UTAPHY-HEP-2}
\rightline{June,16,1992}
\vskip 6pt

\begin{center}

IMPLICATIONS OF NEW GALLEX RESULTS FOR THE MSW SOLUTION \\
OF THE SOLAR NEUTRINO PROBLEM

J. M. Gelb \\
{\it NASA/Fermilab Astrophysics Center, \\
     Fermi National Accelerator Laboratory, Batavia, Illinois 60510}

and

Waikwok Kwong and S. P. Rosen \\
{\it Department of Physics, The University of Texas at Arlington, \\
     Arlington, Texas 76019}

\end{center}

\vskip 2ex
\centerline{ABSTRACT}
\begin{quote}
   We compare the implications for \Be7\ and $pp$ neutrinos of the two MSW fits
   to the new GALLEX solar neutrino measurements . Small mixing angle solutions
   tend to suppress the former as electron-neutrinos, but not the latter, and
   large angle solutions tend to reduce both by about a factor of 2. The
   consequences for BOREXINO and similar solar neutrino--electron scattering
   experiments are discussed.
\end{quote}

\vskip 6pt

The impressive results from the GALLEX solar neutrino experiment\Ref{1} clearly
imply that neutrinos from the $pp$ reaction in the sun have been detected, but
the magnitude of the observed signal, $83 \pm 21$ SNU, leaves unresolved the
issue of solar physics versus neutrino physics as the cause of the solar
neutrino problem.\Ref{2}  As pointed out by the collaboration,\Ref{3} it is
possible to stretch solar models to be in conformance with the GALLEX,
Davis,\Ref{4} and Kamiokande~II\,\Ref{5} experiments, and it is also possible
to
fit all of the data with the MSW mechanism.\Ref{6} Here we wish to explore the
implications of the MSW fits for $pp$, \Be7, and \B8\ neutrinos in order to
guide future experiments in the resolution of the basic issue.

In the context of the MSW mechanism, the GALLEX data pick out two small and
distinctive regions of parameter space: one with a small mixing angle ($\sin^2
2\theta = 7 \times 10^{-3}$) and $\Delta m^2 = \hbox{few} \times 10^{-6}$, and
the other with a large mixing angle ($\sin^2 2\theta = 0.6$) and somewhat
larger
$\Delta m^2$. The small mixing angle solution lies on the so-called
`nonadiabatic' line\Ref{7} which is characterised by a falling
electron-neutrino
survival probability as neutrino energy falls. To accommodate the relatively
large GALLEX signal, this fall cannot continue indefinitely into the low energy
regime, but must begin to turn upwards at some energy in the neighborhood of
the
\Be7\ line at 0.86 MeV or between the \Be7\ and $pp$ neutrinos. Such behaviour
occurs in those regions of parameter space in which the adiabatic\Ref{8} and
nonadiabatic solutions join onto one another;\Ref{9} examples of it can be
found
in  Figures (6e and f) of the original paper of Rosen and Gelb.\Ref{10} The
essential point is that while the Davis and Kamiokande~II experiments are
sensitive to the high energy end of the solar neutrino spectrum, the GALLEX
experiment probes the low energy end which might behave in an entirely
different
way.

A similar difference between high and low energy neutrinos shows up in the
large
angle solution.\Ref{11} High energy neutrinos are in the asymptotic regime of
the adiabatic line in which the electron-neutrino survival probability is
given
by\Ref{8} $\sin^2 \theta$, which is roughly 0.2 in the present case, while low
energy neutrinos fall on the other end of the adiabatic probability curve which
gives a larger value of\,\Ref{8} ($1 - 0.5\sin^2 2\theta$), about 0.7 in the
present case. It is important in the large angle case to take into account the
`in vacuo' oscillations undergone by the neutrino on its journey from Sun to
Earth;\Ref{10} this will modify the above limits, bringing them closer
together.

Let us first concentrate on low energy neutrinos and the adiabatic MSW survival
probability\Ref{8} which holds for both the small and large angle GALLEX
solutions.\Ref{3}  With $\Delta m^2$ in the range of $10^{-6}$, the MSW
resonance enhancement occurs deep inside the sun relative to the in vacuo
oscillation length of $\hbox{a few} \times 10^6$ meters for these neutrinos;
and
so we can write the electron-neutrino survival probability at Earth, including
oscillations between Sun and Earth,\Ref{10} as:
\begin{eqnarray}
  P(\nu_e \to \nu_e; \hbox{Earth})  &=& \frac{1}{2} \left( 1 -
           \frac{p(r)\cos^3 2\theta}{\sqrt{(p^2(r) + \sin^2 2\theta}}\right) \\
  p(r) &=& 1.52 \times 10^{-7}\,\frac{E\,\rho(r)}{\Delta m^2} - \cos 2\theta
\end{eqnarray}
where $E$ is the neutrino energy in MeV and $\Delta m^2$ is the squared mass
difference in eV$^2$. The neutrino is produced at radius $r$ and $\rho(r)$ is
the electron density in mol/cc, properly adjusted for Helium and heavy element
abundance.  Since the low energy neutrinos are produced over a region of
roughly
15--20\% of the solar radius around the core,\Ref{2} we must integrate the
survival probability over this region to form:
$$
  \langle P(\nu_e \to \nu_e) \rangle =
                      \int f(r) P(\nu_e \to \nu_e; \hbox{Earth})\,d\,r \eqno(3)
$$
where $f$ is the fraction of neutrinos produced at radius $r$.

The solar density decreases fairly slowly over the production regions for the
\Be7\ and $pp$ neutrinos\Ref{2} and we can infer the qualitative behaviour of
the integrated probability $\langle P(\nu_e \to \nu_e)\rangle$ from the
unintegrated expression in Eq.~(1). For the \Be7\ neutrinos, we fix the energy
at 0.86 MeV and use the tables of solar electron densities and production
fractions as given by Bahcall and Ulrich\Ref{2} to study $P(\nu_e \to \nu_e;
\hbox{Earth})$ as a function of $\Delta m^2$.  In the small angle case, the
survival probability behaves almost like a step function around the enhancement
point $\Delta m^2 = 10^{-5}$ at which $p(r)$ vanishes, changing rapidly from
almost zero below it to close to unity above it. In the large angle case, the
variation is much more gentle: it passes from a value of $0.5(1 - \cos^3
2\theta) = 0.38$ below the enhancement point, now near $\Delta m^2 = 2 \times
10^{-5}$, to $0.5(1 + \cos^4 2\theta) = 0.58$ above it.

The qualitative behaviour of $pp$ neutrinos is exactly the same, except that,
because of the lower energies and a production region that extends further out
into somewhat lower density zones of the sun,\Ref{2} the enhancement points
shift to lower values of $\Delta m^2$. Taking a typical energy of 0.3 MeV, we
find that  enhancement occurs at $3 \times 10^{-6}$ in the small angle case and
$6 \times 10^{-6}$ in the large angle one. In Figures 1 and 2 we show these
features in actual calculations of the integrated probabilities $\langle
P(\nu_e
\to \nu_e)\rangle$ for specific mixing angles. The behaviour of the small angle
curves are not very sensitive to the precise value of the mixing angle,  while
the large angle GALLEX solution\Ref{3} clusters closely around the value used
in
Figure 2.

{}From the curves in Figure 1, we can read off the survival probabilities for
\Be7\ neutrinos in the small angle solution for the $\Delta m^2$ range of
$\hbox{(3 to 10)} \times 10^{-6}$ eV$^2$.  Between 3 and $6 \times 10^{-6}$,
$\langle P \rangle$ remains very small, being less than 0.05 at $6.3 \times
10^{-6}$; it then climbs rapidly to about 0.15 at $8 \times 10^{-6}$ and 0.4 at
$10\times 10^{-6}$. In the same interval the survival probability for $pp$
neutrinos is significantly larger, climbing rapidly from 0.5 to 1.

Curves for the large angle solution, with $\sin^2 2\theta = 0.6$, are shown in
Figure 2. The range for $\Delta m^2$ in this case is $4 \times 10^{-6}$ to $3
\times 10^{-5}$ and the survival probability for \Be7\ neutrinos gradually
varies from 0.38 to 0.56.  The survival probability for $pp$ neutrinos is
slightly  larger over most of the range, increasing from 0.45 to 0.58.

Suppose that we apply these results to experiments which plan to observe \Be7\
neutrinos via neutrino-electron scattering, for example BOREXINO.\Ref{12}  For
recoil electrons in the kinetic energy range of 250 to 663 keV, the lower end
of
which excludes $pp$ neutrino scattering, the ratio of the rate with \Be7\
electron-neutrino survival probability $\langle P \rangle$ at Earth to the rate
in the standard solar model (SSM) is:
$$
      R(\langle P \rangle) = 0.787 \langle P \rangle + 0.213           \eqno(4)
$$
The expected SSM signal in BOREXINO is 47 events per day,\Ref{12} and so the
small angle solution, with $\langle P \rangle$ varying between 0 and 0.4, will
yield between 10 and 25 events per day. The large angle solution yields
$\langle
P \rangle$ in the narrow range of 0.38 to 0.56 and a BOREXINO signal of 24 to
31
events per day. Thus we can conclude that should the BOREXINO signal be
significantly below 25 events per day, say 16, then the small angle solution
will be the correct one. Moreover, the fraction of \Be7\ neutrinos arriving at
Earth as electron-neutrinos will be much smaller than the average fraction for
\B8\ neutrinos, which stands at 40\% on the basis of the Kamiokande~II
experiment, and this would rule out a small change in solar temperature\Ref{13}
as the cause of the solar neutrino problem.

Should the BOREXINO signal fall in the neighborhood of 25 events per day, then
the situation will be ambiguous with respect to the two solutions, and we will
have either to measure the $pp$ neutrinos directly or to measure the spectral
shape of the high energy neutrinos in order to choose between them. In the
small
angle solution, $pp$ neutrinos remain almost entirely as electron-neutrinos,
while in the large angle case the electron-neutrino survival probability is at
most 58\%.  The difference between the two cases should show up in low
temperature experiments designed to detect $pp$ neutrinos via
neutrino--electron
scattering.\Ref{14}

In the case of high energy neutrinos, those with energy greater than 5 MeV, the
small angle and large angle solutions lead to different electron-neutrino
survival probabilities. For small angles, the nonadiabatic approximation yields
a probability of\,\Ref{9,\,15} $\exp(-C/E)$ where the constant $C$ is
approximately  10 MeV;\Ref{10,\,16} whereas for large angles, the adiabatic
approximation (see Eq.~1) gives a constant value at Earth of $0.5(1 - \cos^3
2\theta)$. The difference in these probabilities will be reflected in the
electron recoil spectra in solar neutrino-electron scattering experiments such
as Kamiokande~II,\Ref{5} SNO\Ref{17} and Superkamiokande,\Ref{18} especially in
the  neighborhood of 5 MeV.\Ref{19} In general, measured spectral shapes  which
differ from the SSM predicted shape will also rule out solar physics as  the
cause of the solar neutrino problem.

A BOREXINO signal of about 30 events per day will point to the large angle
solution as being correct. This conclusion would imply that \B8\ neutrinos in
the energy range from 1 to 5 MeV would be roughly 40\% electron-neutrinos. It
would be interesting to observe them directly.

Besides total rate, BOREXINO also has the recoil electron spectrum as a
diagnostic tool: should it observe a shape significantly different from that
predicted by SSM, then the solution to the solar neutrino problem  must lie in
neutrino physics rather than solar physics. The same holds true if  a day-night
effect,\Ref{20} which favors the large angle solution, were to be  observed.
The
SSM spectrum and representative cases of the spectra for the MSW  mechanism are
shown in Figure 3.

In conclusion, we see that experiments designed to detect \Be7\ neutrinos, such
as BOREXINO, have a good chance of resolving the basic question of the solar
neutrino problem and of distinguishing between MSW solutions. As more data
accumulates, the GALLEX errors\Ref{1} should decrease and thus help to remove
those areas of parameter space\Ref{3} leading to ambiguous predictions for
\Be7\
neutrinos.

The authors wish to thank W. Hampel, T. Kirsten, J. Bahcall, and J. Weneser for
useful conversations.

The research of J. M. Gelb is supported in part by the U. S. Department of
Energy and by NASA grant NAGW-2381; the research of S. P. Rosen is supported in
part by U. S. Department of Energy grant DE-FG05-92ER40691.

\vskip 4ex

\parskip 0ex
\setlength{\parindent}{-1em}
\frenchspacing

\def \jour#1,#2,#3,#4'{{\it #1} {\bf #2}, #3 (19#4)}
%
%  This macro takes four arguments #1 to #4 separated by commas and
%  terminated by an apostrophe ('); They are expanded as shown inside the
%  curly bracket.  E.g. \jour Phys. Lett.,B111,22,92'  will be expanded as
%  {\it Phys. Lett.} {\bf B111}, 22 (1992)
%  Note the use of spaces in the definition and in the result. The following
%  abbreviations can be used for names of journals.
%
\def \PRD{{\it Phys. Rev. D}}
\def \PRL{{\it Phys. Rev. Lett.}}
\def \RMP{{\it Rev. Mod. Phys.}}
\def \PL {{\it Phys. Lett.}}
\def \NP {{\it Nucl. Phys.}}

\def \etal{{\it et al.}}

REFERENCES
\vskip 1ex

\Ref{1}
GALLEX Collaboration preprint GX 1-1992, submitted to {\it Phys. Lett. B},
   June 1, 1992.

\Ref{2}
J. N. Bahcall and R. Ulrich, \jour\RMP,60,297,88'.

\Ref{3}
GALLEX Collaboration preprint GX 2-1992, submitted to {\it Phys. Lett. B},
   June 1, 1992.

\Ref{4}
R. Davis \etal, in {\it Proc. 21th ICRC Adelaide}, edited by R. J. Protheroe,
   Volume 12, 143 (1990).

\Ref{5}
K. Hirata \etal, \jour\PRL,65,1297 and 1301,90'; and \jour\PRD,44,2241,91'.

\Ref{6}
S. P. Mikheyev and A. Y. Smirnov, \jour Nuovo Cim. ,9C,17,86';
L. Wolfenstein, \jour\PRD,17,2369,78'.
A review with references to the  original literature is given by T. K. Kuo and
J. Pantaleone, \jour\RMP,61,937,89'.

\Ref{7}
S. P. Rosen and J. M. Gelb, \jour\PRD,34,969,86';
E. W. Kolb, M. S. Turner and T. P. Walker, \jour\PL,B175,478,86'.

\Ref{8}
H. A. Bethe, \jour\PRL,56,1305,86';
V. Barger, R. J. N.  Phillips and K. Whisnant, \jour\PRD,34,980,86'; and
A. Messiah in {\it '86  Massive Neutrinos in Physics and Astrophysics},
edited by O. Fackler and J. Tran  Than Van (Editions Frontieres, Paris 1986)
p.373.

\Ref{9}
W. C. Haxton, \jour\PRL,57,1271,86'.

\Ref{10}
see Rosen and Gelb, Reference 7.

\Ref{11}
S. J. Parke and T. P. Walker, \jour\PRL,57,2322,86'.

\Ref{12}
R. S. Raghavan \etal, \jour\PRD,44,3786,91'.

\Ref{13}
J. N. Bahcall, {\it Neutrino Astrophysics} (Cambridge University Press, 1989);
S. A. Bludman, D. C. Kennedy, and P. G. Langacker, \jour\NP,B374,373,92'.

\Ref{14}
B. Cabrera, L. M. Krauss, and F. Wilczek, \jour\PRL,55,25,85';
L. M. Krauss and F. Wilczek, \jour ibid,55,122,85';
R. E. Lanou, H. J.  Maris, and G. M. Seidel, \jour ibid,58,2498,87';
S. R. Bandler \etal, \jour ibid,68,2429,92'.
These papers also discuss the possibility of observing \Be7\ neutrinos
simultaneously with the $pp$ ones.

\Ref{15}
S. J. Parke, \jour\PRL,57,1275,86'.

\Ref{16}
J. N. Bahcall and H. A. Bethe, \jour\PRL,65,2233,90';
H. A. Bethe and J. N. Bahcall, \jour\PRD,44,2962,91'.

\Ref{17}
A. B. McDonald in {\it Franklin Symposium in Celebration of the Discovery of
the
Neutrino}, Philadelphia, April 30 1992 (to be published).

\Ref{18}
Y. Totsuka in {\it Franklin Symposium in Celebration of the Discovery of the
Neutrino}, Philadelphia, April 30 1992 (to be published).

\Ref{19}
W. Kwong and S. P. Rosen, \jour\PRL,68,748,92'.

\Ref{20}
A. J. Baltz and J. Weneser, \jour\PRD,35,528,87' and D{\bf 37}, 3364 (1988);
M. Cribier \etal, \jour\PL,B182,89,86';
E. D. Carlson, \jour\PRD,34,1454,86';
A. Dar \etal, \jour ibid,35,3607,87'.

\nonfrenchspacing

\pagebreak

\parskip 2ex

FIGURE CAPTIONS

Figure 1.~~Electron-neutrino survival probability as a function of $\Delta m^2$
in the small angle solution. The solid curve, marked $E_{\nu} = 0.862$ MeV, is
for the mono-energetic \Be7\ branch of the solar neutrino spectrum, and the
dashed curve is for a typical $pp$ neutrino with energy = 0.300 MeV. The mixing
angle is taken to be $\sin^2 2\theta = 7 \times 10^{-3}$, but the curves are
not
sensitive to small changes in the range allowed by Reference 3.

Figure 2.~~Electron-neutrino survival probability as a function of $\Delta m^2$
for the large angle solution $\sin^2 2\theta = 0.6$. The solid curve with
energy
$E_{\nu}$ = 0.862 MeV is for \Be7\ neutrinos and the dashed curve, with
$E_{\nu}$ = 0.300 MeV is for a typical $pp$ neutrino.

Figure 3.~~Recoil electron spectra for the scattering of \Be7\ neutrinos by
electrons. The solid curve is for the Standard solar model, the dashed one is
for a typical large angle solution, and the dash-dotted curve is for a typical
small angle solution. The actual shapes are not very different from one
another, but the overall normalisations are significantly different.

\end{document}